\documentclass[12pt,preprint]{aastex}

\newcommand{\invcm}{cm$^{-1}$}
\newcommand{\kms}{km\,s$^{-1}$}
\newcommand{\mic}{$\mu \mathrm m$}
\newcommand{\mucep}{$\mu \,\mathrm{Cep}$}

\slugcomment{In preparation for ApJ, 2006}

\shorttitle{The 12 \mic\ spectrum of $\mu$ Cep}
\shortauthors{Ryde et al.}

\begin{document}


\title{Water vapor on supergiants.\\
The 12\,\mic\ TEXES spectra of $\mu$ Cephei}


\author{N. Ryde}
\affil{Department of Astronomy and Space Physics, Uppsala University, Box 515, SE-75120 Uppsala, Sweden}
\email{ryde@astro.uu.se}

\author{M. J. Richter\altaffilmark{1}}
\affil{Department of Physics, University of California at Davis, CA 95616}
\email{richter@physics.ucdavis.edu}

\author{G. M. Harper}
\affil{Center for Astrophysics and Space Astronomy - Astrophysics Research Lab, 593 UCB, University of Colorado, Boulder, CO 80309-0593}
\email{gmh@casa.colorado.edu}

\author{K. Eriksson}
\affil{Department of Astronomy and Space Physics, Uppsala University, Box 515, SE-75120 Uppsala, Sweden}
\email{kjell.eriksson@astro.uu.se}

\author{D. L. Lambert}
\affil{Department of Astronomy, University of Texas at Austin, RLM 15.308, USA, TX 78712}
\email{dll@astro.as.utexas.edu}


\altaffiltext{1}{Visiting Astronomer at the Infrared Telescope Facility,
which is operated by the University of Hawaii under Cooperative Agreement
no. NCC 5-538 with the National Aeronautics and Space Administration, Office
of Space Science, Planetary Astronomy Program.}


\begin{abstract}
Several recent papers have argued for warm, semi-detached, molecular
layers surrounding red giant and supergiant stars, a concept known as a MOLsphere.
Spectroscopic and interferometric analyses have often corroborated this general
picture.
Here, we present high-resolution spectroscopic
data of pure rotational lines of water vapor at $12\,$\mic\ for the supergiant $\mu$ Cep.
This star
has often been used to test the concept of molecular layers around supergiants.
Given the prediction of an isothermal, optically thick water-vapor layer in Local Thermodynamic Equilibrium around the star (MOLsphere),
we expected
the $12\,$\mic\ lines to be in emission or at least in absorption but filled in by emission from the
molecular layer around the star. Our data, however, show the contrary;
we find definite absorption.
Thus, our data do not
easily fit into the suggested isothermal MOLsphere scenario. The
$12\,$\mic\ lines, therefore, put new, strong constraints on the MOLsphere concept and on
the nature of water
seen in signatures across the spectra of early M supergiants.
We also find that the absorption is even stronger
than that calculated from a standard, spherically symmetric model photosphere without any
surrounding layers.
A cool model photosphere, representing cool outer layers is, however, able to reproduce the
lines, but this model
does not account for water vapor emission at 6 \mic. Thus, a unified model for water
vapor on \mucep\
appears to be lacking.
It does seem necessary to model the underlying photospheres of these supergiants
in their whole complexity.
The strong water vapor lines clearly reveal inadequacies of classical model
atmospheres.
\end{abstract}





\section{Introduction}
\label{intro}

Water is a key molecule in the
atmospheres of oxygen-rich stars of spectral type late M and cooler.
It is one of the most abundant molecules
and it is a dominant
source of opacity in the infrared. Therefore, it plays an important
role both for the photospheric structure and
the appearance of infrared spectra. To different degrees it dominates the
spectra of brown dwarfs, M-dwarfs and Mira variable stars.
Water lines and bands seem to be a common feature of M
supergiants and M giants  \citep[see for
example][]{tsuji_1997,tsuji_1998,yam_99,tsuji_ny,tsuji_2000,tsuji:iau:01,verhoelst}.
While hydrostatic and homogeneous models of
stellar atmospheres do not predict water vapor spectral features for stars
hotter than late M-type, the existence of water vapor in mid-K to
early-M giants
was observed from medium-resolution ($R=\lambda/\Delta\lambda\approx1600$),
near-infrared spectra recorded
by the \emph{Infrared Space
Observatory (ISO)} \citep{tsuji_1997,tsuji_2000, tsuji_2001}.
These features are
attributed to a static, warm, molecular layer
situated within a few stellar radii beyond, and dynamically detached from, the photosphere - an empirical concept
known as a MOLsphere\citep[see][]{tsuji_1997, tsuji_1998,tsuji_2000,matsuura} -
but distinct from the cool, expanding
circumstellar envelope. The water vapor, at temperatures of $1000$--$2000\,\mbox{K}$
\citep{tsuji_1997}, results in non-photospheric signatures in infrared
spectra of M giants. The notion of a molecular layer (MOLsphere)
has been used in several cases
to explain discrepancies between synthetic spectra and
near-infrared observations of late K and M giants and supergiants
showing water-vapor features.

Further, mid-infrared spectra with a spectral resolution of $R\approx 80,000$
revealed water vapor signatures in even warmer stars,
such as Arcturus, a K2IIIp giant \citep{ryde:02,ryde:03}. This is the warmest
star yet with water vapor features in its disk-averaged
spectrum, and it is indeed a surprising discovery.
Thus, the presence of water vapor in
the spectra of K and M stars is now well established \citep{tsuji:03}.

In their analysis of the water lines in Arcturus, \citet{ryde:02, ryde:livermore}
argued that the water
vapor is photospheric and not circumstellar. The line
widths of the water vapor lines match those of the OH lines that
are definitely formed in the photosphere \citep[see the discussion in][]{ryde:02}.  No velocity
shifts between the water and the OH lines are found.  The authors were
able to model both the water vapor lines and the OH lines of
different excitation by one
photospheric model after cooling the temperature structure of
the outermost layers to lower than that calculated by a  standard
model photosphere. The later model requires that the flux is conserved
through every layer of the modeled star; at the
cooler temperatures, water vapor forms.
A further discussion on possible
causes for the presence of the water vapor lines is also given.
Thus, \cite{ryde:02} suggest that these lines are
formed in the photosphere of the star and that, in order to model these lines, hydrostatic,
homogeneous model photospheres assuming
\emph{Local Thermodynamic Equilibrium} (LTE) are inadequate.
Standard models of the outer photospheres
of giants and supergiants omit potentially important physics, such
as inhomogeneities that
result from large convective motions and the break-down of the
assumption of LTE in the calculation of the physical structure of the photosphere.

A challenge to the current understanding of the MOLspheres
surrounding M supergiants, are the high-resolution,
mid-infrared spectra at 12 \mic\ of the M1-M2 Ia-Iab supergiant Betelgeuse presented by \citet{ryde:06_aori}.
They show that neither published MOLsphere models, introduced
to account for other observations of Betelgeuse, nor
a synthetic spectrum calculated on the basis of a classical model photosphere at
the often assumed effective temperature for Betelgeuse, match their spectra.
However,
with the uncertainties in the outer regions of the photospheric models
in mind, they are able to model the lines
with a cooler temperature structure in the
line-forming regions. This is the same sort of explanation as they invoked for the existence of water vapor in the spectrum of
Arcturus, a star that has no evidence of a MOLsphere. Thus, they argue that it is not necessary to introduce a MOLsphere in order to
reasonably explain the 12 \mic\ lines in the spectrum of Betelgeuse.
It should be noted that no
definitive solution is presented as to the origin of the $12\,$\mic\ lines that also can
explain water-vapor signatures at other wavelengths in the spectrum of Betelgeuse.

To summarize, the location -- whether photospheric or
circumstellar -- and the origin of the water vapor signatures found in a range of K and M giants and
supergiants are not clearly and consistently established.  However, it is clear that a growing
body of spectroscopic data for a range of red giants and supergiants suggests that
classical model atmospheres are inadequate \citep{tsuji:03}. It is unclear whether
the water
vapor features have a common origin across the range of stars and wavelengths
observed.
In this present paper, we present observations of
the M1-M2 Ia supergiant $\mu$ Cep, a prototypical MOLsphere case.

\section{Evidence for water-vapor in $\mu$ Cep}

\begin{figure}
\epsscale{1.00}
\plotone{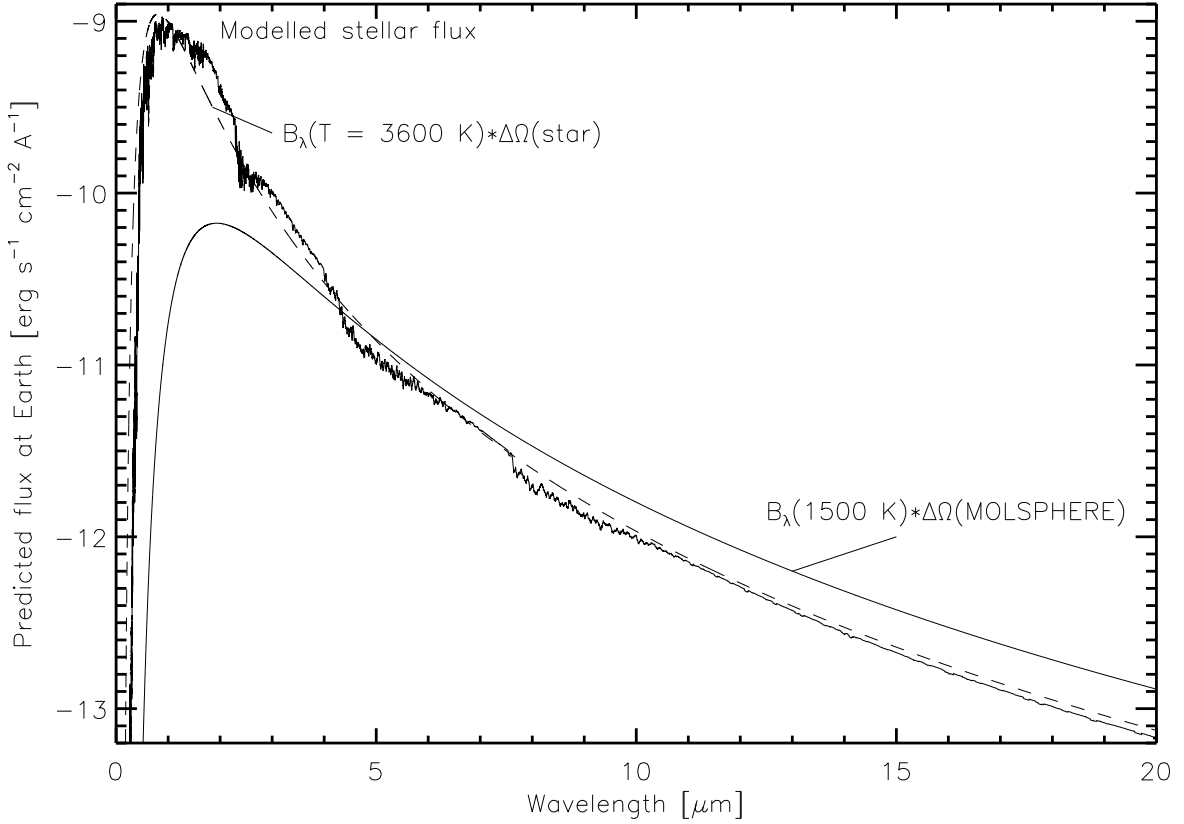}
\caption{Model Spectral Energy Distribution of \mucep\ is plotted together with
a flux from the corresponding Planck function of the stellar effective temperature and assuming a radius of the star of
650 R$_\odot$ and a distance to it of 390 pc \citep{perrin:05}. Furthermore, the flux from the
isothermal MOLsphere is plotted
as calculated from a Planck function of 1500 K and assuming a radius of the MOLsphere of approximately two stellar radii. For optically thick lines this gives the flux in the lines. From the figure it is evident that
optically thick lines are expected to be in absorption for wavelengths shorter than approximately
$5\,$\mic\ and in emission for longer wavelengths,
in accordance with \citet{tsuji_ny, tsuji:03}.
 \label{sed}}
\end{figure}

\citet{danielson:65} suggested that spectral signatures in absorption of vibrational-rotational bands
at 1.4 \mic\ (mostly $\nu_1\nu_2\nu_3 = 101-000$ transitions) and 1.9 \mic\ ($ \nu_1\nu_2\nu_3  = 011-000$) in low-resolution observations of \mucep\ made with the \emph{Stratoscope II} telescope
were due to water vapor. This suggestion of stellar water vapor was, however, overlooked
until it was confirmed by \citet{tsuji_2000}, \textbf{who in \citet{tsuji_ny} also} found evidence for water vapor absorption at
$2.7\,$\mic\ ($\nu_1$ and $\nu_3$ fundamental vibrational-rotational
bands). Furthermore, \emph{ISO} data analysed by \citet{tsuji_ny}
show evidence of water vapor in emission at 6.3 \mic\ ($\nu_2$ fundamental vibrational-rotational
band) and at 40 \mic\ (pure rotational lines).
Thus, the 1.4, 1.9, and 2.7 $\mu$m features are seen in absorption, whereas the 6.3 and
40 $\mu$m features are detected in emission. \citet{tsuji_ny,tsuji:03} shows, in an intuitive interpretation of the observed data,  that this
can be achieved and explained elegantly in an optically thick MOLsphere consisting of water vapor. For wavelengths shorter than
$5\,$\mic\ this layer causes absorption while at longer wavelengths
the geometrical extension of the water vapor sphere leads to
emission, see also Figure \ref{sed}.
This layer is
approximately one stellar radius beyond the photosphere, has an excitation
temperature of 1500 K and a column density
of $N_\mathrm{col}\mathrm{(H_2O)}=3\times 10^{20}$\,cm$^{-2}$ \citep{tsuji_2000,tsuji:03}.
The temperature of this layer is intermediate between the hot chromosphere
and the cold circumstellar matter. The water molecules in the layer are not thought to be of photospheric origin, but rather
formed at a distance beyond the photosphere and absent beneath the layer. \citet{perrin:05}
corroborate the idea of molecular
layers around $\mu$ Cep, based on
an analysis and modeling of narrow-band spatial interferometry in the K band.


The first authors to publish observations of stellar $12\,$\mic\ lines of water vapor were,
to our knowledge, \cite{antares} in
spectra of Betelgeuse and Antares at a resolution of $R\approx8000$.
The 12 $\mu$m pure rotational
lines have previously not been observed for the supergiant \mucep, and the interesting question arose as to how these water lines
would behave, whether they are in absorption or in emission. 

If the MOLsphere scenario with its optically thick water vapor lines is correct, then these
lines ought to be in emission, like the 6.3 and 40 $\mu$m
features, assuming that the lines are optically thick, see Fig. \ref{sed}.
In a photospheric scenario without a major MOLsphere, they would more likely be
in absorption. An obvious third possibility is a mixed
case of contributions from two components, i.e. a variation of the MOLsphere in which it contributes optically thin emission that
weakens the photospheric absorption lines.

\begin{figure}
\epsscale{1.00}
\plotone{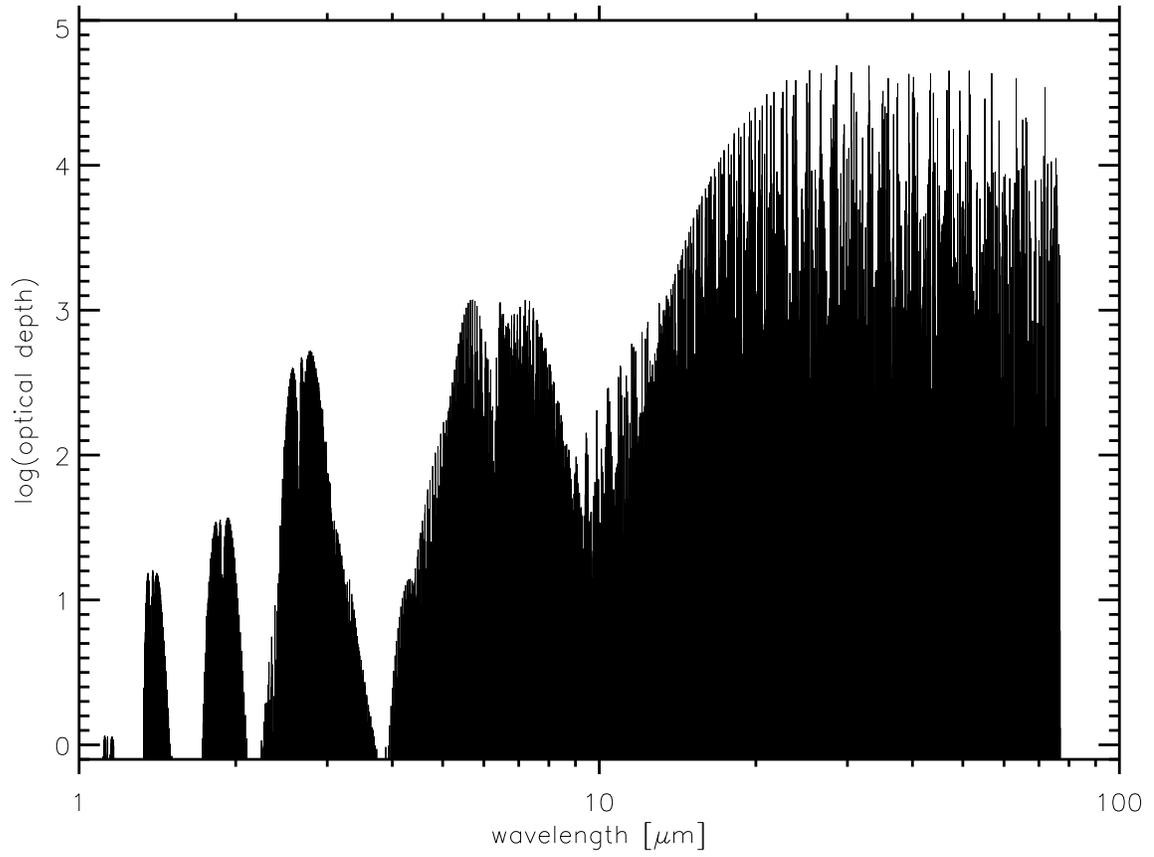}
\caption{The logarithmic optical depth of water-vapor lines from 1 to 77 \mic\
in the isothermal MOLsphere scenario. The temperature is assumed to be 1500 K and the
column density $3\times10^{20}\,\mathrm{cm^{-2}}$ \citep{tsuji_2000,tsuji:03}.
 \label{tau}}
\end{figure}

In order to investigate the optical depths of water vapor lines from the isothermal MOLsphere, we
calculate their optical depths from line data taken from the water vapor line-list
of \citet{par} and the model
parameters in Tsuji (2002b, 2003).
At line center, $\nu_0$, assuming a thermally broadened line with a characteristic line width of
$\Delta\nu_\mathrm D = \nu_0/c \times (2kT/m_\mathrm{H_2O})^{1/2}$ we get
\begin{equation}
\label{equ}
\tau^l_{\nu_0} \approx 0.02654\times \frac{N_\mathrm{col}}{\sqrt{\pi}\Delta\nu_\mathrm D}\, \,(1-e^{-h\nu_0/kT})\,\frac{g_l f_{lu}}{U(1500\,\mathrm{K})}\,e^{-\chi/kT},
\end{equation}

where $N_\mathrm{col}$ is the column density of water vapor, $U$ is the partition function of water vapor and has a value of 2713 for a temperature of 1500 K \citep{par}, and $\chi$ is the excitation energy
of the level.  In Figure \ref{tau} we plot this optical depth for water vapor from $1$ to $77\,$\mic.
From the figure we can immediately
see that there are optically thick water vapor lines in all wavelength bands, in agreement
with the conclusion of \citet{tsuji_2000,tsuji:03}. In particular, many $12\,$\mic\ lines
are optically very thick. As an illustration, from Eq. \ref{equ} and using line data presented in Table \ref{water}, we
see that $\tau^l_{\nu_0} \approx 200$ for
the water line at $818\,$\invcm.


\section{Observations and data reduction}


Our observations of $\mu$ Cep were made with TEXES, the Texas
Echelon-cross-Echelle Spectrograph (Lacy et al. 2002), at the \emph{IRTF}.
The observations were made at the start of the night
of Dec 2, 2000 (UT) to confirm the telescope focus and registration of
the telescope acquisition camera with the TEXES entrance slit for subsequent
science targets.

We observed in the
TEXES high spectral resolution mode.
The spectral resolution, as determined by
Gaussian fits to telluric atmospheric lines, has a FWHM equivalent
to $R=65 000$ or 4.6 \kms.
The slit width was 1.5\arcsec\ and the length 8\arcsec.
The cross-dispersed spectrum included 9 separate orders covering the
spectral range from 813.59 \invcm\ to 819.48 \invcm.  We used
telluric lines to determine that our wavelength scale is accurate to
$0.4 \pm 0.6$ \kms.
At this setting,
the individual spectral orders are larger than the 256$^2$ pixel
detector, resulting in slight gaps in the final, extracted spectrum.

Two data files were saved to disk.  Each file consists of 16 nod-pairs
where we nodded 10\arcsec\ North-South to take the target completely off the slit.
We integrated for 2 seconds
on either side of the nod.  On later TEXES observation runs, we found
this pattern to be quite inefficient and now integrate 6-8 seconds before
nodding the telescope.  Total integration time for $\mu$ Cep is 64
seconds.

Before each data file, we observed a calibration sequence consisting of
an ambient temperature blackbody, a low emissivity surface, and
blank sky.  Subtracting sky from the blackbody provides a telluric
absorption spectrum suitable for establishing the wavelength scale.
Furthermore, the black-sky difference frame gives an approximate correction
for telluric interference; if the sky, telescope, and instrument optics
were all at a single uniform temperature, then dividing by the black-sky
difference would, in theory, perfectly correct for telluric absorption.

We reduced the data according to standard procedures
described in~\citet{texes}.
The spectra were extracted using the spatial profile along the slit
for each file.  Data from the two files were then combined with
weighting according to the effective integration time (the square
of the signal-to-noise ratio).

In each order we fitted a 5$^{th}$ order polynomial to normalize the continuum.
Spectral regions with
stellar and telluric features were given low weight in the fit.
We expect any systematic uncertainties in the polynomial fits to
reduce the equivalent widths determined by roughly 10\%\ \citep{ryde:06_aori}.

\section{The observed spectra of \mucep}

\begin{figure}
\epsscale{1.00}
\plotone{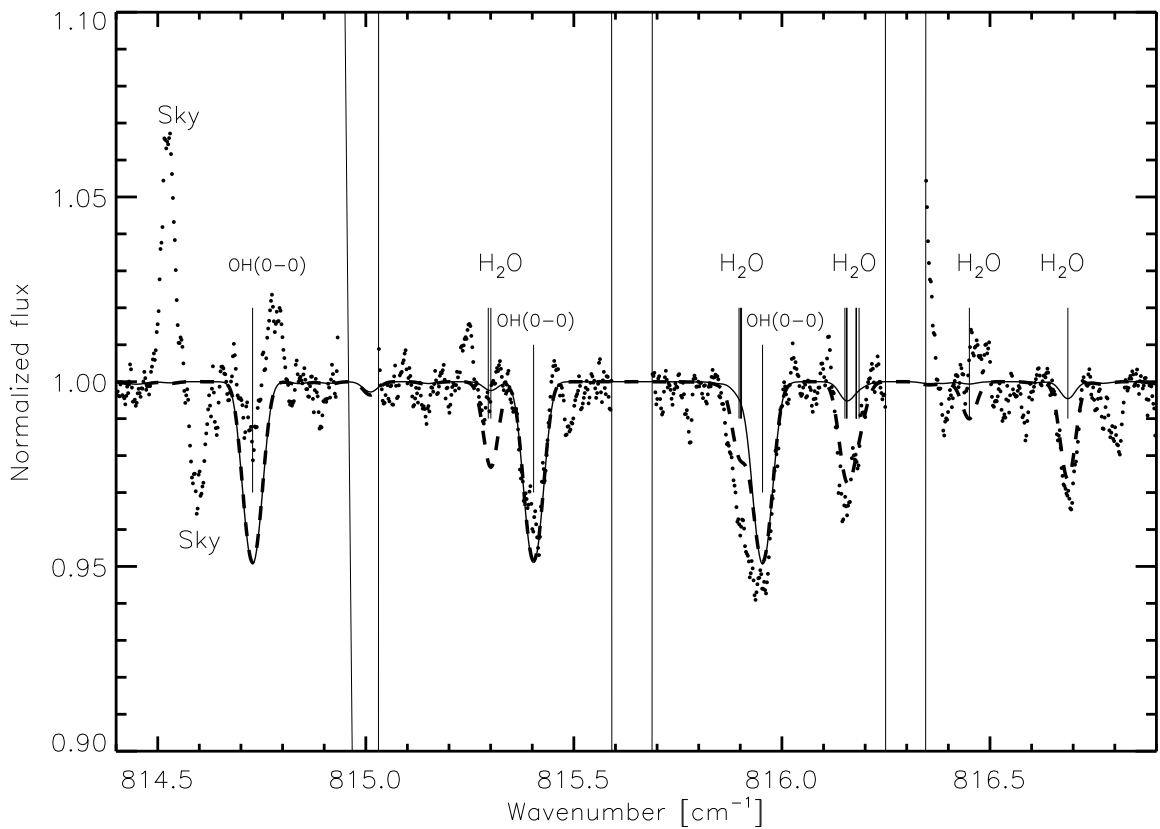}
\caption{Recorded normalized flux spectrum of $\mu$ Cep is shown with dots. The wavelength range spans
$12.23-12.28$\, \mic. Our synthetic
spectrum, based on a classical model photosphere with an effective temperature of 3600 K,
combined with an expected optically thin dust continuum, is shown by the full line together with
the observed spectrum. A synthetic spectrum based on a temperature of 3250 K is shown by a
dashed line. The orders, which are approximately $0.5\,\mathrm{cm^{-1}}$ wide are marked
with the vertical lines. The most important lines are identified and marked.
 \label{fig1}}
\end{figure}

\begin{figure}
\epsscale{1.00}
\plotone{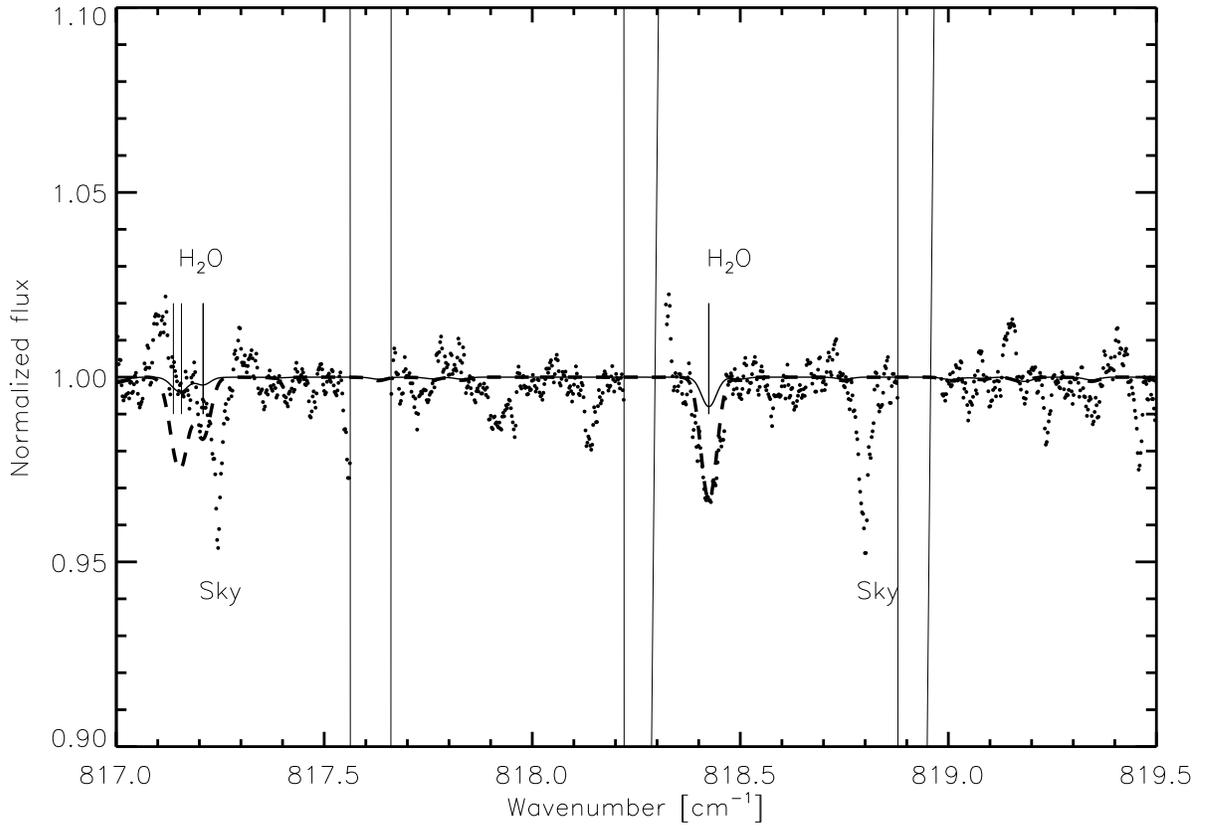}
\caption{The continuation of the spectrum, see legend of Figure \ref{fig1}.
 \label{fig2}}
\end{figure}

In Figures \ref{fig1} and \ref{fig2} we show our recorded spectrum of $\mu$ Cep, shifted
to the laboratory frame.
The orders of the spectrometer are shown by the vertical lines.
\textbf{Four} water line features are clearly detected, namely those at \textbf{$815.90$\,\invcm\ (a pure rotational line
within the first excited
vibrational state, $\nu_2$)}, $816.15-816.19$\,\invcm\ (also pure rotational lines within the first excited
vibrational state, $\nu_2$), 816.69 \,\invcm\ (\textbf{a rotational line
within the} ground state), and the two lines at $818.42$\,\invcm\ (\textbf{also rotational lines
within the ground state}).
Two OH lines of a quartet are also identified. Three of these lie within the wavelength range observed. One of these (814.7 \mic) and
the water lines at 815.3, 816.45, and 817.15 \mic\ have not been identified, most probably because of problems with sky subtraction and
noise, and the difficulties in the continuum subtraction.
In Tables \ref{OH} and \ref{water} we give the observational data measured from these spectra.
Although we have measured only very few lines, the lines are found to be shifted
by the same amount, namely $\langle\Delta
v_\mathrm{OH}\rangle$ = $23.9\pm
1.4\,\mbox{km\,s$^{-1}$}$
and
$\langle\Delta
v_\mathrm{water}\rangle$ = $21.3 \pm
1.1\,\mbox{km\,s$^{-1}$}$ in the frame of the observer.
\clearpage
\begin{table*}
  \caption{Observational data of pure rotational, OH lines.
} \label{OH}
  \begin{center}
  \begin{tabular}{lccccccc} \hline
  \noalign{\smallskip}
 \multicolumn{1}{c}{$\tilde{\nu}_\mathrm{lab}$}     &
  \multicolumn{1}{c}{$E''_\mathrm{exc}$} &
  \multicolumn{1}{c}{$\log gf$} &
  \multicolumn{1}{c}{$v'-v''$}    &
  \multicolumn{1}{c}{Lower level} &
  \multicolumn{1}{c}{$\tilde{\nu}_\mathrm{lab}-\tilde{\nu}_\mathrm{obs}$} &
  \multicolumn{1}{c}{Obs. Equivalent} &
  \multicolumn{1}{c}{FWHM\tablenotemark{a}}
  \\
   \multicolumn{1}{c}{ } &
   \multicolumn{1}{c}{ } &
   \multicolumn{1}{c}{ } &
   \multicolumn{1}{c}{ } &
   \multicolumn{1}{c}{ }&
   \multicolumn{1}{c}{} &
   \multicolumn{1}{c}{Width\tablenotemark{a}\tablenotetext{a}{A colon (:) marks measured values with large
uncertainties.}} &
   \multicolumn{1}{c}{ }
     \\
   \multicolumn{1}{c}{[cm$^{-1}$]} &
   \multicolumn{1}{c}{[eV] } &
   \multicolumn{1}{c}{ } &
   \multicolumn{1}{c}{ } &
   \multicolumn{1}{c}{ } &
   \multicolumn{1}{c}{[$10^{-3}$\,cm$^{-1}$]} &
   \multicolumn{1}{c}{[$10^{-3}$\,cm$^{-1}$]} &
   \multicolumn{1}{c}{[km\,s$^{-1}$] }

  \\
 \noalign{\smallskip}
  \hline
 \noalign{\smallskip}
814.7280 & 1.297 & -1.57 & 0-0 & R$_\mathrm{1e}24.5$ &  -  & - & - \\ 
815.4032 & 1.302 & -1.58 & 0-0 & R$_\mathrm{2e}23.5$  &  63 & 2.5 & 22\\ 
815.9535  & 1.300 & -1.57 & 0-0 & R$_\mathrm{1f}24.5$ &  68 & 4.5: & 26:\\ 
\noalign{\smallskip}
  \hline
  \end{tabular}
  \end{center}
  \smallskip
\end{table*}

\begin{table*}
  \caption{Observational data and the line list of the water-vapor lines.
}
\label{water}
  \begin{center}
  \begin{tabular}{lccccccc} \hline
  \noalign{\smallskip}
 \multicolumn{1}{c}{$\tilde{\nu}_\mathrm{lab}$}     &
   \multicolumn{1}{c}{$E''_\mathrm{exc}$} &
  \multicolumn{1}{c}{$\log gf$} &
  \multicolumn{1}{c}{$J'$$K'_a$$K'_c$$J''$$K''_a$$K''_c$}    &
  \multicolumn{1}{c}{$v_1v_2v_3$}&
  \multicolumn{1}{c}{$\tilde{\nu}_\mathrm{lab}-\tilde{\nu}_\mathrm{obs}$} &
  \multicolumn{1}{c}{Equivalent} &
  \multicolumn{1}{c}{FWHM}
  \\
   \multicolumn{1}{c}{ } &
  \multicolumn{1}{c}{} &
   \multicolumn{1}{c}{ } &
   \multicolumn{1}{c}{ } &
   \multicolumn{1}{c}{ } &
   \multicolumn{1}{c}{} &
   \multicolumn{1}{c}{width}
    \\
   \multicolumn{1}{c}{[cm$^{-1}$]} &
   \multicolumn{1}{c}{[eV] } &
   \multicolumn{1}{c}{ } &
   \multicolumn{1}{c}{ } &
   \multicolumn{1}{c}{ } &
   \multicolumn{1}{c}{[$10^{-3}$\,cm$^{-1}$]} &
   \multicolumn{1}{c}{[$10^{-3}$\,cm$^{-1}$]} &
   \multicolumn{1}{c}{[km\,s$^{-1}$] }
  \\
 \noalign{\smallskip}
  \hline
 \noalign{\smallskip}
816.157    & 1.467 & -1.109  &     & &   \\
816.151    & 1.467 & -1.586  &       &  &   \\
816.15525  & 1.150 & -1.301  & 22    13 10   21  12   9  & (010)& 57 & 1.7 & 17   \\
816.179    & 2.721 & -1.241  &       &&    \\
816.179    & 2.721 & -0.764  &      & &   \\
816.18551  & 1.150 & -1.778  & 22    13   9    21  12   10 & (010)&    \\
816.68703  & 1.014 & -1.35  & 24    12   13  23  11   12 & (000) & 62 & 1.5 & 15 \\
818.4238\tablenotemark{c}\tablenotetext{c}{Assignments from Tsuji (2000b).}
           & 1.029 & -1.66  & 22    16  6    21  15    7 & (000)&    \\
818.4247\tablenotemark{c}    & 1.029  & -1.19  & 22    16   7    21  15    6 & (000) & 56 & 1.8 & 21 \\
\noalign{\smallskip}
  \hline
  \end{tabular}
  \end{center}
  \smallskip

\end{table*}
\clearpage

\section{The modelling of the spectra of $\mu$ Cep}

We have modeled the photospheric spectrum at 12 \mic\ of $\mu$ Cep and a spectrum generated by a MOLsphere surrounding the
star. $\mu$ Cep is not identical to Betelgeuse but is a supergiant resembling Betelgeuse
\citep{tsuji_2000, verhoelst_ke}.
Therefore, we have used our photospheric model of Betelgeuse, as presented in Ryde et al. (2006) as
a typical supergiant model to which our data will be compared.
We assume that this model is a reasonable representation of the photosphere of $\mu$ Cep.

The model photosphere we used was calculated with
the {\sc marcs} code \citep{marcs:03}. These hydrostatic, spherical model photospheres are
computed assuming LTE,
chemical equilibrium, homogeneity, and the
conservation of the total flux (radiative plus convective; the
convective flux being computed using the mixing length
recipe). The  radiation field used in the model generation is
calculated with absorption from atoms and molecules by opacity
sampling at approximately 95\,000
wavelength points over the wavelength range $1300\,\mbox{\AA} $--$
20\,\mbox{$\mu$m}$. The models are calculated with 56 depth points from
a Rosseland optical depth of $\log \tau_\mathrm{Ross}=2.0$ out to $\log
\tau_\mathrm{Ross}=-5.0$, which in our case corresponds to an
optical depth evaluated at $500\,\mbox{nm}$ of $\log
\tau_{500}=-4.1$. The physical height above the $\log
\tau_\mathrm{Ross}=0$ layer of this outermost point is
$2.8\times10^{12}\,\mbox{cm}$ or $6\%$ of the stellar radius.
Extending the model out to $\tau_\mathrm{Ross}=-5.6$  affects the
water-vapor lines only marginally.

Our stellar parameters, used as input for the model photosphere calculation are
$T_\mathrm{eff}=3600$~K, $\log g= 0.0$, solar metallicity (except for the C, N, and O abundances), $M=15\,M_\odot$
(implying $R=650\,$R$_\odot$), $\xi_\mathrm{micro}=4\,$\kms\ (microturbulence) and
$\xi_\mathrm{macro}=7\,$\kms\ (macroturbulence). As a comparison \cite{levesque} deduce
an effective temperature of 3700 K for $\mu$ Cep and find a $\log g= -0.5$.
The C, N, and O abundances are
$A_\mathrm C = 8.29$\footnote{The abundance by number $N_\mathrm C$ is given through the following definition:
$A_\mathrm C\equiv \log N_\mathrm C - \log N_\mathrm H + 12$}, $A_\mathrm N = 8.37$, and $A_\mathrm O = 8.52$.
The decreased carbon abundance and the increased nitrogen abundance compared to
solar values are consistent with signatures of the expected
first dredge-up which massive stars like Betelgeuse and $\mu$ Cep experience. The signatures are those of CN-cycled material
which affects the surface abundances since the convective envelope of the star during the first dredge-up
reaches down to nuclear-processed regions of the interior \citep{lambert:84}.
The most abundant molecules in the deep atmosphere are molecular hydrogen, CO, OH, CN, and CH. Above the layer
defined by $\log \tau_\mathrm{Rosseland}=-1$, water starts to form and its abundance becomes greater than that of CN and CH.

The isothermal, spherical MOLsphere, assumed to be in LTE, is characterized by an inner and an outer radius,
a single temperature, and a column density of water vapor. The parameters of the MOLsphere of \mucep\ that we have used are
those discussed by \citet{tsuji_ny}, namely $T=1500$~K, $R_\mathrm{in}=2\,\mathrm{R}_\star$, $R_{\mathrm out}= 4\,R_\star$, and
$N_\mathrm{col}=3\times10^{20}\,\mathrm{cm^{-1}}$.
The density in the levitated MOLsphere is assumed to be in hydrostatic equilibrium
in the gravity field of the star \citep{tsuji_ny}, thus in some unknown way being
supported at the inner radius.
The density structure decreases almost exponentially outwards from
the inner boundary with a
scale height which is very much smaller than the shell thickness.
We have further assumed a microturbulence velocity in the MOLsphere of
4 \kms\ and introduced a macroturbulence of $v_{\mathrm D}= 7$ \kms, with which we convolve the synthetic spectrum.
The background flux shining onto the MOLsphere is given by the synthetic spectrum based on our
photosphere ($T_\mathrm{eff}=3600\,$K).

The synthetic spectra of the photosphere and MOLsphere were
calculated for points on the spectrum
separated by  $\Delta \tilde{\nu} \sim 1\,\mbox{km\,s$^{-1}$}$
(corresponding to a resolution of $\tilde{\nu}/\Delta \tilde{\nu}
\sim 300\,000$) even though the final resolution is lower. In the radiative transfer calculation of the MOLsphere component
we use 22 rays on the stellar disk, and 53 shells around the star.
Our models (both the model photosphere and MOLsphere, and the synthetic spectra based on them) are calculated in
spherical geometry. The photosphere has a modeled extension ($\Delta R_\mathrm{atm}/R_\star$) of approximately
5\%. Calculations in spherical symmetry will generally tend to decrease
the strengths of strong lines in the mid- and far-infrared.
In extreme cases even photospheric emission may appear. The reason for this is
that an extended photosphere will occupy a larger solid
angle at line wavelengths where the total opacity is large. This
will result in a larger flux at line centers and therefore weaker
lines than those resulting from calculations in plane-parallel
geometry. A MOLsphere will naturally cause emission in optically thick lines due to its
large geometrical size.

The line list in our calculation of the photospheric synthetic spectra is described in \citet{ryde:02} and is
based on the
work by \citet{poly_1, poly_2}. The wavelengths of the lines in the list are sufficiently accurate
for the resolution of
the observations. However, only the strongest lines are included. For the temperatures in our
model photosphere
only the strongest lines do indeed emerge, thus justifying the use of the line list.
The data of the water lines in the calculation of the
MOLsphere are, however,  taken from
the line-list of \cite{par}, which is more complete than the one used for the photospheric spectrum,
but the wavelengths are not as accurate as those used for the photosphere. For the
cold temperature of the MOLsphere a completeness of the line list is more important than the accuracy of
the wavelengths of the lines. The continuous opacities in the MOLsphere from H$^-$ free-free (ff) and H ff are included in our
calculation.

Following \citet{tsuji:78,tsuji:03} we assume an optically thin dust envelope
with $\tau_{10\,\mu\mathrm m}=0.1$,
that contributes only continuous emission veiling the
photospheric line spectrum. Assuming a distance of approximately $390\pm140$~pc \citep{perrin:05},
the slit corresponds to $(9\pm3)\times 10^{15}\,$cm, which corresponds to approximately 200 stellar radii on the sky.
Contrary to the case of Betelgeuse, which lies approximately three times
closer to us, we here assume that the entire dust envelope is
within the slit and that all dust emission is recorded \citep[cf. the discussion by][]{ryde:06_aori}.
From the \emph{ISO} low-resolution 2-40 \mic\
data and the continuum from a classical model photosphere presented in \citet{tsuji:03},
we have estimated the dust contribution at $12\,$\mic\ to be $f^\mathrm{thin}_\mathrm{dust}=
(\mathrm{Flux_{star+dust}- Flux_{star})/(Flux_{star+dust}}) =77\%$ which is the value we use.
Note, however, that this value is uncertain and will affect the lines strengths directly.

\section{Discussion}

In addition to the OH lines in absorption, we have observed and identified
\textbf{four} clear \emph{absorption}
features due to rotational transitions
of water vapor at 12\,\mic\ from
the supergiant $\mu$ Cep. On the basis of the MOLsphere \textbf{scenario we expected the water lines to be in} \emph{emission}.
The interpretation of these lines
is not straight-forward.

\textbf{The observed absorption lines at $12$\,\mic\ leads us to two separate complications associated with the formation
of the water lines at 12\,\mic. First, the model, proposed in this Section, that fits the spectra implies cooler regions in the
outer photosphere where the lines are formed compared to what is expected from a classical model photosphere.
The line velocities, widths and depths are evidence for a photospheric
origin of the water absorption, but the model photosphere needed for the outer regions does not
fulfill the assumptions of a classical model photosphere. The proposed reasons for the break-down of the classical
assumptions are an inhomogeneous photosphere, or non-LTE conditions. Second, why water-vapor lines show up in emission at $6.3$ and $40$\,\mic\ but
not at $12$\,\mic\ is unexplained, but this fact should
contain a clue to the atmospheric environment of \mucep\ and further modelling and discussion
of these features is needed. It can neither be
explained by a simple photosphere
with a cooler outer structure, since that predicts all water lines across the spectrum to be in
absorption, nor can it be explained by an optically thick, isothermal MOLsphere.
Therefore, it would be of great interest to try to understand the 12 \mic\ spectra combined with all other
spectroscopic
water vapor data and interferometric data, and try to explain them all in terms of a consistent,
unified picture, since
$\mu$ Cep is a prototypical MOLsphere case \citep{tsuji_ny}.
A consistent scenario which can account for the water-vapor features and the nature of water in supergiants is
lacking. Some sort of non-classical atmosphere is clearly needed, but the challenge to explain all features
in one model still prevails.}

\subsection{A cooler outer photosphere?}

In Figures \ref{fig1} and \ref{fig2} we have also plotted our
model spectra based on a photosphere of an effective temperature of 3600 K, the assumed temperature of \mucep. The spectrum includes
the contribution of the optically thin dust emission.
While we see from the figures that the OH quartet, unfortunately represented by only one clean line,
is quite well modeled with our combined photospheric and optically-thin dust model,
the water vapor lines show stronger absorption than
one would expect from a classical model photosphere. Changing the effective
temperature of the model photosphere by $\pm 100$ K affects the equivalent widths of the
water-vapor lines by  approximately $\pm 50\%$,
and an increase of $\log g$ by 0.3 dex increases the lines' strengths by approximately 40\%, far from what is
needed to model the observed absorption lines.

The water vapor lines
and OH lines both depend on the partial pressure of
oxygen in the same way. 
However, the lines of these two molecules react differently to a change in
temperature.
In order to fit the water lines, a photospheric model of an effective temperature of 3250 K is needed.
A synthetic spectrum based on such a cold model
is plotted in the figures as well. The fit is very good for both the OH and the various water vapor lines.
It should be noted, however, that there is not a very large range in excitation energies of the water lines
($1.014 - 1.150$ eV). Ideally, a test
of a model structure would include lines of different excitations.
The cooler model of 3250 K was also used for the similar case of
Betelgeuse in order to model 12\,\mic\ spectra \citep{ryde:06_aori}.
These authors discuss the details of this model,
such as the differences in C, N, and O abundances between the models.

An effective temperature of 3250 K is out of the range of possible
temperatures for
\mucep.
As mentioned in Section \ref{intro}, a way of interpreting the cold temperature structure that fits the 12\,\mic\ lines, is that classical models
fail to describe relevant physical features such as inhomogeneities,
structures not in LTE, etc in supergiants.
The colder structure could be representing cold convectional
elements in an inhomogeneous photosphere. Alternatively, the cooler model could be simulating a colder outer
photosphere in the regions where the water-vapor lines are formed.
It would, therefore, be interesting to investigate the influence of inhomogeneities due to convection on the water-vapor lines and
the detailed effects on the spectra of a change of the outermost atmospheric structure, which is known to be
uncertain. We will address these questions in a forthcoming paper.

The turbulent line-widths of the 12\,\mic\ lines and the OH lines (which are resolved at $R=65000$ or $4.6\,$\kms) are of the same order
in both \mucep\ and Betelgeuse.  Apart from the different amounts of circumstellar dust, these
two supergiants seem to behave in the same way concerning the $12\,$\mic\ lines.
As was discussed in \citet{ryde:06_aori} the line widths in Betelgeuse
indicated a photospheric origin.
Furthermore, there is no significant velocity
shift between the OH and the water vapor lines.
This finding is similar to
what we found for other stars we have analysed; Arcturus (K1.5III, Ryde et al. 2002) and Betelgeuse
\citep[M1-2 Ia-Iab,][]{ryde:06_aori}.
Thus, based on their line widths and velocities, the observed 12\,\mic\ lines in \mucep\
look like photospheric absorption lines.
As a note, if the 12\,\mic\ lines were formed in a
MOLsphere, this water layer would need to have
large turbulent velocities and no expansion velocity.

To summarize, a photosphere with a non-classical outer structure is able to describe the
mid-infrared spectra, both the OH and water vapor lines. Furthermore, the lack of velocity shifts and
the line widths are nicely explained. However, this explanation does not explain other infrared
and interferometric data \citep{perrin:05}. Thus, the simple photosphere
with a cooler outer structure can not explain all observed features since that model predicts
all water lines across the spectrum to be in absorption.

\subsection{A modified MOLsphere?}

\begin{figure}
\epsscale{1.00}
\plotone{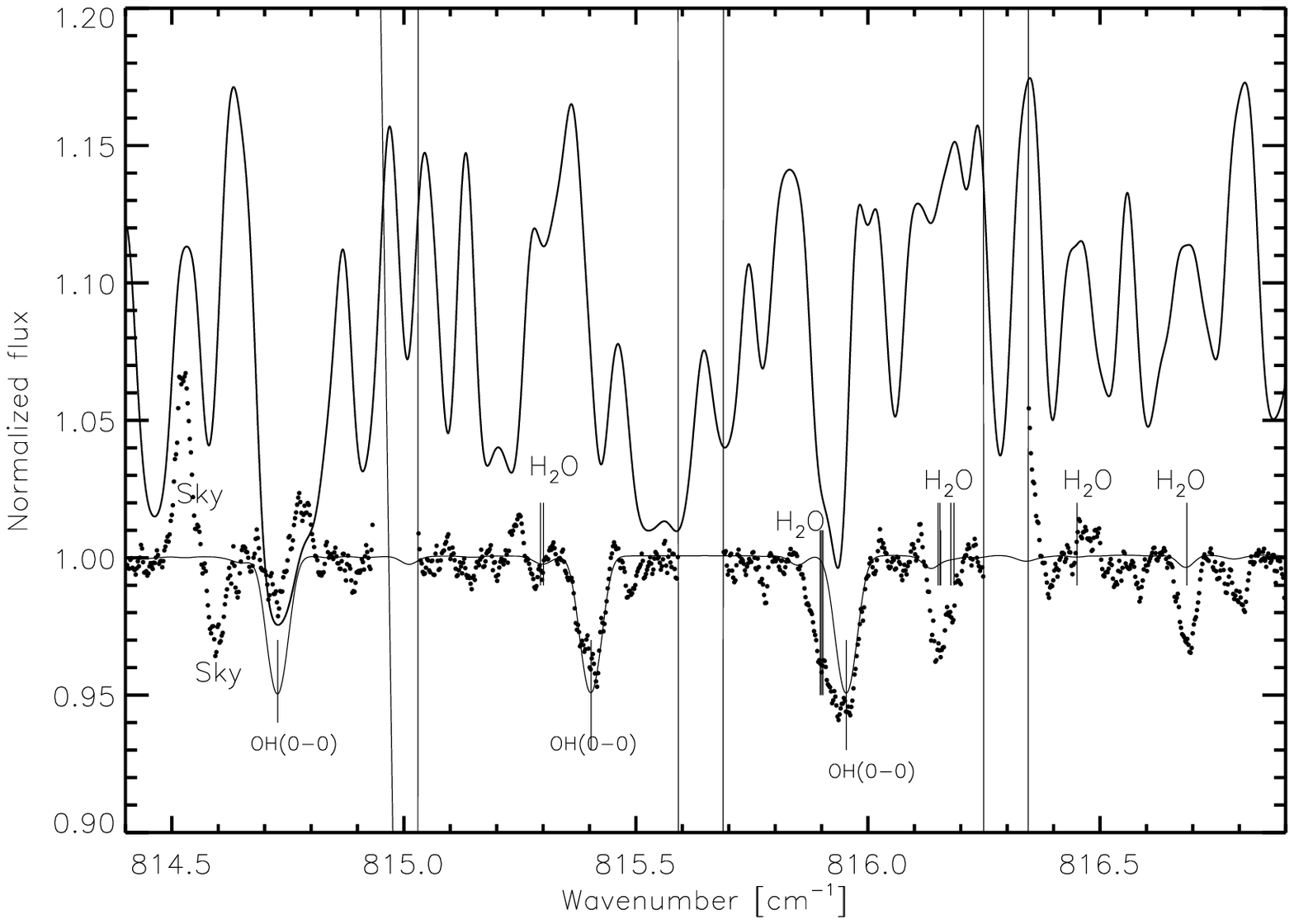}
\caption{The normalized, observed spectrum of \mucep\ from 12.24 -12.28\,\mic\ is
shown by dots. The full line with a continuum at 1.0 is a photospheric spectrum calculated
with the \citet{par} line list. The MOLsphere results shown as the emission spectrum by the full line
above the observations are calculated with
$T=1500$~K, $R_\mathrm{in}=2\,\mathrm{R}_\star$, $R_{\mathrm out}= 4\,R_\star$, and
$N_\mathrm{col}=3\times10^{20}\,\mathrm{cm^{-1}}$. This spectrum is plotted on the same ordinate scale
as the other two spectra. We see that the large, optically thick MOLsphere results in huge emission features
which are not seen in the observations. The continuum in the MOLsphere is due to free-free H and H$^{-}$ emission,
but this continuous opacity is very small.
This leads to an optically
very thin continuum in the MOLsphere which means that it is transparent in the continuum.
One of the photospheric absorption lines
of OH is not fully filled in by water vapor emission. The other two are. Note, that this MOLsphere realization
does not include OH lines in the calculation.
 \label{molsphere}}
\end{figure}

As we have noted, earlier investigations of \mucep\ have shown that vibrational-rotational water-vapor bands in the near-infrared
are formed in absorption while
the vibrational-rotational $\nu_2$ band at 6.3 \mic\ and the rotational lines at 40 \mic\ are observed in
emission \citep{tsuji_2000}. This behavior
is explained with an optically thick, isothermal MOLsphere
consisting of water vapor \citep{tsuji_2000}.
To directly fit into
this MOLsphere concept with optically thick lines, the
rotational lines at 12 \mic\ are expected to be in emission too. We now see the opposite.
Figure \ref{molsphere} shows the result of
a calculation of the MOLsphere for \mucep, which clearly does not explain
the observed spectrum.

How can we modify the MOLsphere model so that it better predicts the 12 \mic\ lines? We have not solved this
problem, but we discuss a few suggestions here.

\begin{itemize}
\item From Figure \ref{sed} we see directly that, assuming that the lines are optically
thick, it would be possible to
construct a MOLsphere that would give absorption lines at
12 \mic\ by changing the size of the modeled molecular layer. This effect is also explained nicely
by \citet{ohnaka:04a}.
This change would, however, at the same time
lead to the 6 \mic\ lines
also being in absorption, which is not observed \citep{tsuji_ny}.

Despite the assumed similarities between the supergiants \mucep\ and Betelgeuse, the nature of the observed water vapor
is somewhat different, which is discussed by \citet{tsuji_ny}. Contrary to the case of \mucep,  the \emph{ISO} spectrum of Betelgeuse
shows the $\nu_2$ water band at $6.3\,$\mic\ in absorption \citep{tsuji_ny}.
However, the rotational lines  at 40\,\mic\ in \emph{ISO} spectra of
Betelgeuse could also be interpreted as emission, similar to what is seen in  the spectra of \mucep\ \citep{tsuji_ny}.
The mid-infrared spectra
of Betelgeuse show rotational lines at 12\,\mic\ in absorption \citep{antares, ryde:06_aori}, similar to our
\mucep\ observations.
Thus, the two stars show some differences in the nature
of the water assumed to be surrounding them. \cite{tsuji_ny} suggests that the inner boundary of the
surrounding assumed water-vapor envelope is closer to the star in the case of Betelgeuse, which
could explain that the 6.3 \mic\ data are in absorption and not the 40 \mic\ features. However, the MOLsphere suggested
by \cite{tsuji_ny} does not fit the high-resolution spectra at 12 \mic\ either \citep{ryde:06_aori}.

\item The MOLsphere model is based on the assumption that the relative strengths of the different
water lines and bands are given by the excitation temperature of $T_\mathrm{ex}=1500$~K through a Boltzmann distribution affecting their opacities. This may, however, be
questionable. For example, \citet{hinkle_lambert:75} show, for diatomic molecules, that even in the outer regions of a photosphere of a supergiant like $\alpha$ Ori, collisions
are not dominant over radiative processes for vibration-rotational transitions. In the case of water-vapor, the relevant rotational transitions have very strong transition probabilities,
even stronger than the probabilities of relevant vibration-rotation lines. This means that radiation may well be the dominant
transition process, leading to scattering by water-vapor lines.
If  scattering in the optically-thick lines is allowed for, absorption may
be the result. All line photons will be scattered and eventually a fraction of these will be absorbed by the medium,
implying that fewer photons escape. However, this process has to differentiate between the long-wavelength rotational lines which are
seen in emission and the rotational lines at 12\,\mic\ which are seen in absorption. Also, light from the surrounding dust,
which shines strongly at 12 \mic, will
be scattered by the 12\,\mic\ lines and has to be incorporated. Thus, a full non-LTE analysis of a water vapor layer in the
radiation field of the star and the dust would have to be performed. This possibility should be tested in
future MOLsphere models.

If light resonantly scattered in molecular lines is important, special care has to be taken when
comparing spectra from various instruments with different beam sizes. For example, while the slit
dimensions of the SWS spectrometer onboard \emph{ISO} is $14\arcsec \times20\arcsec$ at 6 \mic, the
slit of the TEXES spectrometer is $8\arcsec \times1.5\arcsec$. More scattered light may therefore be
recorded by
ISO than by TEXES. This could be an \textbf{explanation} to why \emph{ISO} measured emission at
6 \mic\ and TEXES absorption at 12 \mic.

\item Another way of modifying the isothermal MOLsphere interpretation is to allow for temperature
gradients. This could be tuned to give absorption lines, but should be physically reasonable. Again, the
emission at 6 and 40\,\mic\ \textbf{also} has to explained.

\end{itemize}

Thus, it would be interesting to investigate different implementations and modifications of a MOLsphere lying beyond
a calculated photosphere of a generic M supergiant (representing the cases of $\mu$ Cep and Betelgeuse)
or a semiempirical model photosphere, and to test these against all available data.


\section{Conclusions}

We have pointed to the accumulating
amount of independent evidence for the existence of water vapor in K and M giants and supergiants.
Explaining existing water lines, bands, and interferometric data of supergiants appears, however, to be a challenge.
In order to understand the origin of
the unexpected water vapor, it seems that we have to go beyond classical modelling, be it a MOLsphere scenario,
models relaxing the assumptions of homogeneity and LTE, or both. No consistent
unified model, which can without modifications account for all available data, exists today.

\mucep\ is a test bench for the MOLsphere
concept, introduced to explain a range of unexplained observational features, both interferometric and spectroscopic, albeit
at low spectral resolution.
The new high-resolution, spectroscopic water-vapor 12 \mic\ data presented here do not fit into
the current MOLsphere model that can explain other infrared and interferometric data of $\mu$ Cep. In a non-modified, optically-thick version,
the 12 \mic\ lines were expected to be in emission, but the opposite is found. Furthermore, a pure photospheric model predicts much too weak absorption lines for the effective temperature expected for the supergiant.
We show that a photosphere with a cooler temperature structure in the line-forming regions
of the $12\,$\mic\ lines is able to fit the spectra, possibly providing a clue to their origin.
This indicates that a cooler structure due to, for instance, convective motions, can explain
the recorded water-vapour and OH lines. A problem is that this model does not account for the emission at 6 \mic.

Thus, a consistent picture is still lacking. It is time for more elaborate modelling and to acquire more high-quality data,
especially high-resolution infrared spectra of, for example, \textbf{the 1.4, 1.9, 2.7, 6.3, and 40 \mic\ }wavelength regions. Furthermore,
measurements with the Atacama Large Millimeter Array (ALMA) at frequencies of 100-700 GHz will be able to resolve scales
at which the MOLsphere should show up. Interferometric observations of the angular diameter as a function of frequency will
reveal the nature of the MOLspheres, their existence and their density structure. Note, however,  that for \mucep\
the spatial scales of ALMA are not as well matched
as for Betelgeuse, a star also assumed to have a MOLsphere.
Future studies will have to show whether a modified MOLsphere scenario in addition
to  more sophisticated
photospheric models are able to simulate all spectroscopic and interferometric data available.
Our new data need to be incorporated in any model or scenario of the
atmospheres of $\mu$ Cep in particular and of
supergiants in general.

We have once again shown that pure rotational lines of water vapor in the
mid-infrared, observed with high-resolution spectrographs, are
interesting diagnostics of the atmospheres of (super-)giant stars.
The usefulness of the lines is connected with their relatively large transition probabilities and their location in an
uncrowded part of the spectrum.










\acknowledgments

We would like to thank Bengt Gustafsson and John H. Lacy for fruitful suggestions and
enlightening discussions. We acknowledge the support we received from John H. Lacy, Daniel T. Jaffe, and
Qingfeng Zhu of the TEXES team during the
observations of $\mu$ Cep.
We are grateful for the help of the {\sc irtf} staff.
This work was
supported in part by the Swedish Research Councils ({\sc vr} and {\sc stint}),
the Robert A. Welch
Foundation of Houston, Texas,
{\sc nsf} grant AST-0307497 and NASA grant NNG04GG92G.
Observations with TEXES were
supported by {\sc NSF} grant AST-0205518.


\begin{thebibliography}{}

\bibitem[\protect\astroncite{{Danielson} et~al.}{1965}]{danielson:65}
{Danielson}, R.~E., {Woolf}, N.~J., and {Gaustad}, J.~E., 1965,
\newblock {ApJ} {141}, 116

\bibitem[\protect\astroncite{{Gustafsson} et~al.}{2003}]{marcs:03}
{Gustafsson}, B., {Edvardsson}, B., {Eriksson}, K., {J{\o}rgensen}, U.~G.,
  {Mizuno-Wiedner}, M., and {Plez}, B., 2003,
\newblock in {IAU Symposium 210: 'Modelling of Stellar Atmospheres´, Eds. N.
  Piskunov, W. W. Weiss, and D. F. Gray}, pp CD--A4

\bibitem[\protect\astroncite{{Hinkle} and {Lambert}}{1975}]{hinkle_lambert:75}
{Hinkle}, K.~H. and {Lambert}, D.~L., 1975,
\newblock {MNRAS} {170}, 447

\bibitem[\protect\astroncite{{Jennings} and {Sada}}{1998}]{antares}
{Jennings}, D.~E. and {Sada}, P.~V., 1998,
\newblock {Science} {279}, 844

\bibitem[\protect\astroncite{{Lacy} et~al.}{2002}]{texes}
{Lacy}, J.~H., {Richter}, M.~J., {Greathouse}, T.~K., {Jaffe}, D.~T., and
  {Zhu}, Q., 2002,
\newblock {PASP} {114}, 153

\bibitem[\protect\astroncite{{Lambert} et~al.}{1984}]{lambert:84}
{Lambert}, D.~L., {Brown}, J.~A., {Hinkle}, K.~H., and {Johnson}, H.~R., 1984,
\newblock {ApJ} {284}, 223

\bibitem[\protect\astroncite{{Levesque} et~al.}{2005}]{levesque}
{Levesque}, E.~M., {Massey}, P., {Olsen}, K.~A.~G., {Plez}, B., {Josselin}, E.,
  {Maeder}, A., and {Meynet}, G., 2005,
\newblock {ApJ} {628}, 973

\bibitem[\protect\astroncite{{Matsuura} et~al.}{1999}]{matsuura}
{Matsuura}, M., {Yamamura}, I., {Murakami}, H., {Freund}, M.~M., and {Tanaka},
  M., 1999,
\newblock {A\&A} {348}, 579

\bibitem[\protect\astroncite{{Ohnaka}}{2004}]{ohnaka:04a}
{Ohnaka}, K., 2004,
\newblock {A\&A} {421}, 1149

\bibitem[\protect\astroncite{{Partridge} and {Schwenke}}{1997}]{par}
{Partridge}, H. and {Schwenke}, D., 1997,
\newblock {J. Chem. Phys.} {106}, 4618

\bibitem[\protect\astroncite{{Perrin} et~al.}{2005}]{perrin:05}
{Perrin}, G., {Ridgway}, S.~T., {Verhoelst}, T., {Schuller}, P.~A., {Coud{\' e}
  Du Foresto}, V., {Traub}, W.~A., {Millan-Gabet}, R., and {Lacasse}, M.~G.,
  2005,
\newblock {A\&A} {436}, 317

\bibitem[\protect\astroncite{{Polyansky} et~al.}{1996}]{poly_1}
{Polyansky}, O.~L., {Busler}, J.~R., {Guo}, B., {Zhang}, K., and {Bernath},
  P.~F., 1996,
\newblock {J. Mol. Spectrosc.} {176}, 305

\bibitem[\protect\astroncite{{Polyansky} et~al.}{1997}]{poly_2}
{Polyansky}, O.~L., {Tennyson}, J., and {Bernath}, P.~F., 1997,
\newblock {J. Mol. Spectrosc.} {186}, 213

\bibitem[\protect\astroncite{{Ryde} et~al.}{2006}]{ryde:06_aori}
{Ryde}, N., {Harper}, G.~M., {Richter}, M.~J., {Greathouse}, T.~K., and {Lacy},
  J.~H., 2006,
\newblock {ApJ} {637}, 1040

\bibitem[\protect\astroncite{{Ryde} et~al.}{2003a}]{ryde:03}
{Ryde}, N., {Lacy}, J.~H., {Richter}, M.~J., {Lambert}, D.~L., and
  {Greathouse}, T.~K., 2003a,
\newblock in {ASSL Vol. 283: Mass-Losing Pulsating Stars and their
  Circumstellar Matter}, p. 227

\bibitem[\protect\astroncite{{Ryde} et~al.}{2002}]{ryde:02}
{Ryde}, N., {Lambert}, D.~L., {Richter}, M.~J., and {Lacy}, J.~H., 2002,
\newblock {ApJ} {580}, 447

\bibitem[\protect\astroncite{{Ryde} et~al.}{2003b}]{ryde:livermore}
{Ryde}, N., {Lambert}, D.~L., {Richter}, M.~J., {Lacy}, J.~H., and
  {Greathouse}, T.~K., 2003b,
\newblock in {ASP Conf. Ser. 293: 3D Stellar Evolution}, p. 214

\bibitem[\protect\astroncite{{Tsuji}}{1978}]{tsuji:78}
{Tsuji}, T., 1978,
\newblock {A\&A} {68}, L23

\bibitem[\protect\astroncite{{Tsuji}}{2000a}]{tsuji_ny}
{Tsuji}, T., 2000a,
\newblock {ApJ Lett.} {540}, L99

\bibitem[\protect\astroncite{{Tsuji}}{2000b}]{tsuji_2000}
{Tsuji}, T., 2000b,
\newblock {ApJ} {538}, 801

\bibitem[\protect\astroncite{{Tsuji}}{2001a}]{tsuji:iau:01}
{Tsuji}, T., 2001a,
\newblock in {IAU Symposium 205: 'Galaxies and their Constituents at the
  Highest Angular Resolutions´, Ed. R. T. Schilizzi}, p. 316

\bibitem[\protect\astroncite{{Tsuji}}{2001b}]{tsuji_2001}
{Tsuji}, T., 2001b,
\newblock {A\&A} {376}, L1

\bibitem[\protect\astroncite{{Tsuji}}{2003}]{tsuji:03}
{Tsuji}, T., 2003,
\newblock in {ESA SP-511: Exploiting the ISO Data Archive. Infrared Astronomy
  in the Internet Age}, p.~93

\bibitem[\protect\astroncite{{Tsuji} et~al.}{1997}]{tsuji_1997}
{Tsuji}, T., {Ohnaka}, K., {Aoki}, W., and {Yamamura}, I., 1997,
\newblock {A\&A} {320}, L1

\bibitem[\protect\astroncite{{Tsuji} et~al.}{1998}]{tsuji_1998}
{Tsuji}, T., {Ohnaka}, K., {Aoki}, W., and {Yamamura}, I., 1998,
\newblock {Ap\&SS} {255}, 293

\bibitem[\protect\astroncite{{Verhoelst} et~al.}{2006}]{verhoelst_ke}
{Verhoelst}, T., {Decin}, L., {Van Malderen}, R., {Hony}, S., {Cami}, J.,
  et~al., 2006,
\newblock {A\&A} {447}, 311

\bibitem[\protect\astroncite{{Verhoelst} et~al.}{2003}]{verhoelst}
{Verhoelst}, T., {Decin}, L., {Vandenbussche}, B., {Van Malderen}, R., and
  {Waelkens}, C., 2003,
\newblock in {IAU Symposium 210: 'Modelling of Stellar Atmospheres´, Eds. N.
  Piskunov, W. W. Weiss, and D. F. Gray}, pp CD--E14

\bibitem[\protect\astroncite{{Yamamura} et~al.}{1999}]{yam_99}
{Yamamura}, I., {de Jong}, T., and {Cami}, J., 1999,
\newblock {A\&A} {348}, L55

\end{thebibliography}

\clearpage

\end{document}